# Concevoir des formations facilitant l'émergence de nouvelles significations face à des évènements inédits et critiques

## Simon Flandin

Université de Genève – Simon.Flandin@unige.ch

## Deli Salini

IUFFP Lugano – Deli.Salini@iuffp.swiss

## Artemis Drakos

Université de Genève – artemis.drakos@gmail.com

## Germain Poizat

Université de Genève – Germain.poizat@unige.ch

Notre recherche s'inscrit dans un programme de recherche technologique en formation des adultes conduit en référence au programme du « cours d'action » (Theureau, 2006, 2009, 2015). À partir de l'hypothèse d'activité-signe inscrite dans ce programme, nous concevons les formations comme des occasions de perturbation et/ou relance de la dynamique de signification des participant-es. Cette contribution a pour buts (i) de faire avancer la conceptualisation de formations conçues comme des aides aux participant-es pour la compréhension et la transformation de leur activité, et (ii) d'élaborer des principes de conception transversaux qui permettent la mise en œuvre de ces formations dans différents contextes. Nous nous appuyons sur l'analyse de formations visant soit la gestion, soit le dépassement d'évènements vécus comme inédits et critiques par les individus concernés. Il s'agit, selon le cas, de formations ayant une visée « réparatrice » de situations vécues comme traumatisantes et/ou sclérosantes, ou ayant une visée « préparatrice » à des situations imprévues et pouvant avoir des conséquences majeures. À travers le concept d'abduction, les conditions d'émergence de nouvelles significations sont décrites, soit à des fins « réparatrices » (dénouement de situations d'impasse), soit à des fins « préparatrices » (préfiguration de situations de crise). Les principes de conception de ces formations et leurs effets sont analysés à l'aide des outils conceptuels du cours d'action, en intégrant deux dimensions supplémentaires : fictionnelle et événementielle. La contribution au programme du cours d'action et à la recherche en formation est discutée en fin d'article au prisme de l'hypothèse d'activité-signe.

formation, activité-signe, abduction, impasse, programme de recherche technologique

*Training design fostering the emergence of new meanings toward unprecedented and critical events*

Our research is part of a technological research program in adult education conducted in reference to the "course of action" program. Using the activity-sign hypothesis of this programme, training situations are thought as opportunities to perturb/relaunch the participants'dynamics of meaning. Our contribution aims to (i) improve the conceptualization of training situations that are thought as aids to the understanding and transformation of the participants situations, and (ii) derive cross-cutting design principles that allow the enactment of these training situations in different contexts. We rely on the analysis of training programs aiming either at the management or at the overcoming of events experienced as unprecedented and critical by the individuals concerned. Depending on the case, these training programs have a "restorative" aim (resolving of impasse situations), or have a "preparatory" aim (prefiguration of crisis situations). The design principles of these training programs and their effects are analysed with the conceptual tools developed within the course-of-action, integrating two additional dimensions: fictional and event-driven. The conditions for the emergence of new meanings are described in terms of abduction, either for reparative purposes (resolution of deadlock situations) or for preparatory purposes (prefiguration of crisis situations). The contribution to course-of-action program and to research on adult education is discussed through the prism of the hypothesis of activity-sign.

adult education, activity-sign, abduction, impasse, design-based research

## 1. Penser la formation dans l'approche du cours d'action

Le programme de recherche cours d'action (CdA) (Theureau, 2006, 2009, 2015) constitue une matrice commune à de nombreuses recherches dans divers domaines de pratique, comme le montre ce dossier spécial. Ces recherches peuvent avoir deux orientations, non exclusives l'une de l'autre : empirique (à enjeux de compréhension) et technologique (à enjeux de conception). Les recherches technologiques ont la particularité de chercher prioritairement à décrire et provoquer des *dynamiques de transformation* de l'activité (potentielles, actuelles, et virtuelles) à des fins de conception d'environnements socio-technico-organisationnels, culturels et éducatifs, quand les recherches empiriques cherchent à documenter et modéliser *des régularités* de l'activité ainsi décrite « pour elle-même ». Ce travail de mise en ordre de l'empirie constitue très souvent une étape pour des recherches technologiques finalisées par la conception de dispositifs, notamment dans le champ de l'éducation et de la formation. Ce faisant, ces dernières ont contribué à valider, enrichir, compléter le programme générique du CdA que ce soit d'un point de vue théorique ou méthodologique (e.g., Poizat, Durand, & Theureau, 2016).

Dans le domaine de la formation des adultes, en particulier au cours de la dernière décennie, le programme de recherche technologique (Durand, 2008 ; Poizat & Durand, 2015) a élaboré des principes de conception de situations (i) de « vidéo-formation », (ii) de simulation en formation, (iii) de formation de formateurs, ou encore (iv) de formation à médiation artistique. Depuis la première systématisation de ce programme proposée par Durand (2008), de nombreuses avancées ont été réalisées. Elles se sont traduites par une agrégation progressive d'hypothèses ontologiques supplémentaires notamment sur *l'individuation-appropriation* (e.g., Poizat & Goudeaux, 2014) ou sur la *mimesis* (e.g., Durand, Goudeaux, Horcik, Salini, Danielian, & Frobert, 2013), par la formulation de différentes hypothèses nontriviales sur la transformation de l'activité (e.g., Durand, 2017), et par une « concrétisation » de plus en plus forte des propositions de conception et des démarches et méthodes associées (Poizat & Flandin, soumis[1]).

Dans ce cadre, la présente contribution s'intéresse aux modalités de construction de significations nouvelles lors, après ou en vue d'événements inédits ou critiques. Cette appréhension de l'inédit et du critique en formation (i) est de prime importance dans divers domaines de pratiques où prédominent par exemple des enjeux de santé (e.g., publics en situations de vulnérabilité) et de sécurité (e.g., industries à risques) et (ii) est centrale dans une perspective de formation « non-curriculaire » (Flandin, Poizat, & Perinet, 2019), à visée développementale, comme pour les formations basées sur la médiation artistique ou la simulation de crises (exemples sur lesquels nous nous appuyons dans cet article).

Nous débutons par une description des apports et conséquences de l'adoption de l'hypothèse d'activité-signe (Theureau, 2006) pour la formation, et des avancées et spéculations autour de la notion d'abduction (Peirce, 1994) ; puis nous conceptualisons sur cette base la notion d'impasse (Salini & Durand, 2020 ; Salini & Poizat, sous presse). Nous nous appuyons ensuite sur l'analyse – parallèle, puis convergente – de deux dispositifs pouvant être considérés comme prototypiques de formations « réparatrice » d'un vécu d'impasse (« le Théâtre du Vécu ») ou « préparatrice » à un vécu d'impasse (un exercice de crise). Suit l'énonciation des principes de conceptions dérivés de cette analyse empirique. La discussion finale porte sur la contribution de l'article à un programme de recherche en formation des adultes et sur les nouvelles perspectives de recherche qui en découlent.

## 1.1. Intérêt et conséquences de l'hypothèse de l'activité-signe dans le domaine de la formation

Les conséquences des hypothèses ontologiques de l'*enaction* et de l'*expérience* (voir l'article introductif à ce dossier spécial) ont largement été abordées dans les synthèses existantes sur le développement du programme CdA en éducation et formation (e.g., Poizat, Salini, & Durand, 2013). L'hypothèse analytique de l'activité-signe a également beaucoup été utilisée dans les recherches en éducation pour décrire et modéliser des phénomènes, dynamiques et trajectoires d'apprentissage-développement (Durand, 2008), d'appropriation, de découverte et de création. En revanche, son « potentiel technologique » reste selon nous insuffisamment exploré, notamment sa fécondité pour comprendre les conditions favorables à des transformations majorantes de l'activité, et pour la conception de situations de formation.

L'hypothèse de l'activité-signe a été formulée par Theureau (2006) à partir de la conjonction de l'hypothèse de l'*enaction* de Maturana et Varela (1994) et de celle de la *pensée-signe* de Peirce (1994), en partant du postulat qu'activité et cognition sont indissociables. Si toute activité est cognitive, et toute cognition est inscrite dans une sémiose – autrement dit dans une dynamique de signification – alors toute dynamique d'activité est une activité-signe (Theureau, 2006). Cette hypothèse permet d'articuler une phénoménologie empirique de l'activité humaine et une théorie généralisée de la signification, et ouvre sur un système de notions descriptives et explicatives qui constitue le cadre d'analyse sémiologique de l'activité humaine (Theureau, 2006, 2009, 2015). Elle porte sur l'expérience vécue de l'acteur et ne sépare pas a priori penser, agir, sentir, ressentir, éprouver, savoir, etc. : les contenus d'expériences sont enactés par l'acteur « de proche en proche », ou plus exactement « de signe en signe ». Si toute activité est à la fois cognitive et signifiante, dans toute situation expériencée par un individu, des dynamiques de transformation des significations sont à l'œuvre et reconfigurent en permanence, et de manière plus ou moins importante, les savoirs et les liens entre les savoirs de l'acteur.

---

[1] En cours d'édition, cet ouvrage constituera un nouveau point d'étape « récapitulatif » du programme.

Ainsi, les « expériences de savoirs » (Dieumegard, 2011) de diverses natures (savoir-faire, connaissances, savoirs symboliques, savoirs non symboliques, etc.) se constituent en cours d'activité et sont considérées comme interprétations inhérentes à la situation que l'acteur fait advenir (Dieumegard, 2009). L'apprentissage est documenté par la construction de *prototypes*, de *types*, et de *relations entre types* (Rosch, 1973). Apprendre consiste pour l'acteur à *typicaliser* ou *typifier*[2] des couplages avec la situation (incluant d'autres acteurs). Ces types peuvent être plus ou moins consensuels au sein d'une communauté et/ou être construits de manière collective à travers une *construction de sens participative*. Ils constituent ensuite des ancrages dans l'appréhension des événements et actions (Schütz, 1987). Un oiseau qu'un individu aperçoit dans une forêt est identifié ou catégorisé dans l'action par des jugements de proximité ou de distance au type ou réseaux de types relatifs à une identité « oiseau » (forme, couleur, son du chant, attributs de la forêt, etc.), qui sont propres à cet individu. Mais ce processus est à l'œuvre quelles que soient les formes d'action. Un physiothérapeute typicalise à chaque instant son expérience et les réactions du patient pendant qu'il réalise une manipulation, et engage des actions sur la base de jugements de typicalité. Dans les situations très complexes, ces types s'articulent et donnent lieu à des configurations (ou reconfigurations) de types également très complexes et difficiles à décrire localement.

Selon cette approche, les apprentissages ne se réalisent donc pas sur un mode séquentiel, par agrégation « mécanique » d'éléments nouveaux dans une base disponible, mais s'expriment par une émergence de nouveauté, sur fond de reconfiguration du déjà-là. Aussi l'activité est-elle, à la fois ou itérativement, conservation et reproduction, mais aussi émergence et transformation qui permettent l'invention, et la création. Précisée plus loin, la notion d'Interprétant, inscrite dans le cadre de l'analyse sémiologique de l'activité élaborée par Theureau (e.g. 2006) à partir de la sémiotique de Peirce, permet de rendre compte précisément de ce principe. Dans cette perspective, la formation poursuit un double objectif : favoriser ce processus de création de la nouveauté sur fond de reconfiguration du présent, et chercher à le rendre « majorant », c'est-à-dire de nature à transformer l'activité de façon souhaitable du point de vue des acteurs et des formateurs. Cette approche n'est donc pas tenue par la spécification « classique » d'une liste d'objectifs pédagogiques précis, communément énoncés de la manière suivante « à l'issue de la formation, les participants seront capables de... ». Le caractère majorant réside dans l'accroissement espéré d'une capacité « globale » des acteurs à éprouver et agir dans des types de situations tendanciellement difficiles. Dans les dispositifs présentés ici, la majoration se concrétise par le dépassement de vécus d'impasse.

## 1.2. Interprétant, signe hexadique et abduction

La notion d'Interprétant (I), mobilisée dans le cadre théorique sémiologique du CdA, s'inspire de celle développée par Peirce qui avait, à travers cette catégorie, inscrit à la fois la mise en œuvre des savoirs acquis et la création de savoirs nouveaux dans le signe. Cette mise en œuvre s'inscrit dans l'ensemble de la sémiose, qui pour Peirce se constitue dans l'interaction triadique[3] entre trois catégories de l'expérience : le « Possible » renvoyant à des propensions qui pourraient ou non s'actualiser dans une situation ; l' « Actuel » renvoyant aux perceptions et réactions significatives et spécifiques du et dans le réel ; le « Virtuel » renvoyant à la généralisation émergeant de l'interaction entre les catégories précédentes, qui infirme ou modifie des significations préalablement constituées. À partir de cette catégorisation, Theureau (2006) élabore la notion de *signe hexadique*[4], une matrice à six composantes dont chacune est reliée dynamiquement aux autres (Tableau 1). Le signe hexadique rend possible la description du *cours d'expérience*, c'est-à-dire un niveau de réduction de l'activité qui prend en considération ce qui est significatif pour l'acteur, montrable, racontable, mimable et commentable par celui-ci, selon des conditions éthiques et méthodologiques particulières.

---

[2] Ce redoublement de vocabulaire est dû à ce que nous rapprochons deux traditions de recherche : l'une issue de la psychologie qui a adopté les termes « typicalisation » et « prototype » (Rosch, 1973), l'autre de la sociologie les termes « typification » et « type » (Schütz, 1987). Nous emploierons désormais les termes « typicalisation » et « type », ce qui simplifie la lecture sans poser, pensons-nous, de problème théorique, compte tenu du niveau de généralité de l'argumentation.

[3] Dans le sens qu'elle n'est pas réductible à des actions entre pairs

Tableau 1 : Les six composantes des trois catégories d'expérience
*Table 1: The six components of the three categories of experience*

| Possible | **E**ngagement (E) : les intentions, préoccupations, états affectifs, ouverts de recherche |
|---|---|
| | **A**nticipation (A) : les attentes ou anticipations et prévisions d'action dans les unités successives |
| | **R**éférentiel (S) : les types, relations entre types et principes d'interprétation appartenant à la culture de l'acteur qu'il peut mobiliser compte tenu de (E) et (A) à un instant donné |
| Actuel | **R**epresentamen (R) : ce qui fait signe pour l'acteur à un instant donné |
| | **U**nité de cours d'expérience (U) : les actions pratiques, discours privés, communications et émotions actualisées |
| Virtuel | **I**nterprétant (I) : les généralisations ou règles d'action ainsi que les manifestations de cycles de recherche (doutes, questions, hypothèses émergeantes) |

La catégorie de l'Interprétant traduit la « transformation » (création, intégration et/ou réorganisation) qui est apportée au Référentiel (donc aux types et relations entre types disponibles pour l'acteur) dans le décours de l'activité. Elle rend manifeste la constante transformation à divers degrés du savoir de l'acteur, de ses habitudes situées (ou de ses typifications), et ce faisant de sa culture propre.

Comprendre ces transformations exprimées par l'Interprétant et surtout comment nous parvenons à connaître quelque chose que nous ne savons pas à partir de quelque chose que nous savons, passe selon Peirce (1868) par la compréhension du rôle des inférences dans la sémiose. Pour cet auteur, nous n'avons pas de pouvoir d'intuition et il nous est donc impossible d'avoir une appréhension immédiate du réel, mais chaque acte de connaissance procède par inférence d'une connaissance précédente. Cette hypothèse n'exclut pas la découverte, mais conduit à l'envisager comme le résultat d'un cas particulier d'inférence : l'abduction. Celle-ci s'ajoute à deux inférences plus connues : la déduction, qui est l'application d'une règle générale à des cas particuliers (qui a ainsi une fonction explicative ou de justification) et l'induction, qui est une dérivation d'une règle depuis un ou plusieurs cas et un résultat (qui a ainsi une fonction de vérification et de généralisation horizontale). L'abduction consiste en l'inférence d'un cas depuis une règle et un résultat, ou plus précisément l'émergence en même temps d'une règle en lien avec un cas, qui lui donne une « raison » (Peirce, 1868).

Avec l'abduction, le degré de nouveauté (tout du moins de nouveauté perçue par l'acteur) est largement supérieur à celui de la déduction ou de l'induction. Activée à partir d'un fait surprenant ou « inattendu » qui contredit des anticipations préalables (Salini, 2013), l'abduction ne génère pas de la certitude, mais fait émerger une hypothèse provisoire (pour l'action), la prédiction d'une possibilité, en tant que choix parmi d'autres possibles, en proposant une nouvelle forme de mise en relation d'éléments inconnus à partir d'éléments connus. Il s'agit d'une hypothèse qui émerge, ou s'enacte, comme tentative de compréhension des « causes possibles » par rapport à des faits observés dont les « vraies causes » sont inconnues. L'abduction s'apparente alors à une forme « d'analogie aventureuse atypique » (Theureau, 2015).

---

[4] Voir également l'article introductif à ce dossier spécial.

Dans les processus abductifs, les dimensions iconiques de la signification en tant qu'hypothèses de compréhension jouent un rôle essentiel (Fisette, 2009 ; Peirce, 1994). Elles s'expriment par l'émergence de signes qui renvoient à leur objet par ressemblance, en s'appuyant sur un ensemble de dimensions corporelles, sensorielles, émotionnelles, ou conceptuelles. Ces signes iconiques (images, diagrammes et métaphores), tout en gardant leur dimension triadique, relèvent davantage du domaine d'expérience du Possible. Ils ont une dimension vague, instable et ouverte, sans contours définis et constituent des hypothèses d'interprétation du monde et en ce sens, des inférences abductives (Fisette, 2009). En particulier, la métaphore propose un nouveau signe, plus précisément une généralisation du lien entre deux signes, fondée sur des similarités « inattendues ». D'une part elle exprime une hypothèse de régularité concernant un lien spécifique entre deux signes, et d'autre part comme toute généralisation (bien qu'hypothétique), elle configure l'anticipation des actions et événements successifs, leur donnant une certaine prévisibilité (Fisette, 2009 ; Salini, 2013).

La modalité de raisonnement abductive est cruciale dans l'émergence de nouvelles significations que ce soit dans des situations ordinaires ou lorsque les acteurs ou les collectifs font face à des évènements, surprenants, inédits, ankylosants et critiques. Peu étudiées en éducation, les inférences abductives sont pourtant déterminantes dans la compréhension de la transformation de l'activité individuelle et collective, mais aussi pour la conception, l'amélioration, ou l'évaluation de certaines situations ou dispositifs de formation (Cunningham, 1998).

La pertinence de l'étude des modalités d'apprentissage-développement en terme de sémiose a déjà été pointée dans la littérature (Bopry, 2007, Colapietro, Midtgerden, & Strand, 2005 ; Midtgarden, 2005, 2010 ; Stables & Gough, 2006) et en particulier l'apprentissage par abduction (Cunningham, 1998 ; Paavola & Hakkarainen, 2005 ; Ventimiglia, 2005). D'après Cunningham, Schreiber et Moss, (2005) l'abduction est souvent négligée pour comprendre l'apprentissage, au profit de l'induction et de la déduction qui prédominent dans la pensée « traditionnelle » de l'enseignement : les enseignants expliquent les règles et informations utiles aux participant-es, qui les mettent à l'épreuve à travers divers exercices et résolutions de problèmes. Dans les dispositifs pédagogiques visant à encourager un raisonnement abductif, les participant-es sont plutôt encouragé-es à observer et interpréter des éléments étonnants et à les interpréter, ce qui encourage (dans le domaine scolaire) la formulation d'hypothèses explicatives et le développement d'un esprit critique (e.g. Ahmed & Parsons, 2013 ; Hwang, Hong, Ye, Wu, Tai, & Kiu, 2019 ; Oh, 2011).

Contrairement aux recherches de ce type considérant que les « enquêtes abductives » (Oh, 2008) ont un intérêt intrinsèque, notre recherche s'intéresse aux conditions (empiriques) dans lesquelles l'abduction joue un rôle développemental déterminant, et à quelles conditions (technologiques) des dispositifs de formation peuvent provoquer et favoriser ce processus. L'objet de cet article n'est donc pas seulement de documenter la construction de nouvelles significations à travers l'abduction, mais aussi de mieux utiliser les processus abductifs pour la conception de situations de formation.

## 1.3. Approche sémiologique de l'« inattendu » et des vécus d'impasse

Dans la dynamique entre connu et inconnu, la notion d'« inattendu » est centrale surtout en considération de la manière dont Peirce (1994) explique la notion de surprise. Celle-ci signale que ce qui est advenu contredit l'attente d'une régularité et fait ainsi émerger la « rupture de l'attente d'une régularité ». Ce faisant elle pointe la possibilité d'une autre régularité. L'« inattendu » est alors quelque chose qui sollicite un changement de nos modes de signification, de nos généralisations constituées auparavant, en activant le mouvement d'investigation et la possibilité de constitution de nouvelles connaissances (Peirce, 1994). Cet « inattendu » peut apparaitre selon différents degrés de surprise, suivant un nuancier de « prévisibilité » et jusqu'à « ce qu'on ne peut pas prévoir » (Grossetti, 2004). On peut ainsi différencier les « inattendus », dont la réalisation est prévisible, mais dont le moment d'occurrence ne l'est pas, et les « impensés », c'est-à-dire des « inattendus » qui n'ont jamais été imaginés et a fortiori décrits et prévus.

Le vécu d'impasse survient lorsque quelque chose entrave la continuité du parcours interprétatif qui émerge lors d'une occurrence inhabituelle, chargée d'un niveau important d'« inattendu » dans l'expérience de l'acteur. En ce cas, une partie de la dynamique globale de signification ne parvient pas à aboutir sur des généralisations établies et ce faisant, elle finit par se replier sur elle-même. Il y a alors émergence d'un questionnement sans réponses qui peut être compulsif et générateur d'affects négatifs. La situation « inattendue » demeure incompréhensible, inexprimable, non partageable, et ce qui est advenu n'est pas dépassé : une gêne, une souffrance ou un sentiment d'inachèvement lui sont associés. Ce vécu d'impasse est souvent renvoyé à des métaphores comme « rue sans issue » renvoyant à quelque chose qui empêche d'aller de l'avant ou de revenir en arrière ; ou de « tourner en rond » rendant compte du fait que les questions sont répétitives, qu'elles restent sans réponse ou reçoivent des réponses peu convaincantes (Salini et Poizat, sous presse). Il s'agit d'une dynamique qui n'ouvre sur rien d'autre qu'elle-même ou qui se resserre, un « nœud » qui paralyse l'activité, car les différents fils s'enchevêtrent sans possibilité apparente de démêlage. Le vécu d'impasse comporte une dimension affective qui est faite à la fois d'inquiétude (jusqu'à l'angoisse) et d'une entrave de la relation avec le collectif (Simondon, 2005). À différents degrés d'intensité, l'individu se sent en détresse, incertain et préoccupé, sinon accablé. Parfois cette détresse se chronicise, et devient un arrière-fond permanent et pernicieux de l'activité. Le sentiment de solitude est marquant et va souvent de pair avec celui de ne pas parvenir à se faire comprendre.

Lors d'un vécu d'impasse, la sémiose – en tant que dynamique inscrite dans une temporalité – est comme court-circuitée : il n'y a plus d'ouverture vers un avenir et un devenir. Les individus se délient de leur processus d'individuation, qui passe par des circuits longs (Stiegler, 2010). Cette déliaison s'effectue soit par une rupture brutale de leur relation au monde (expérience disruptive, sidération), soit par un retour continu à des questions sempiternelles (comme un ressort cassé, le passé est invariablement « présentifié » – Violi, 2014). Cette entrave de la projection vers l'avenir peut être comprise à la lumière de la dimension essentielle de l'expérience de la temporalité et le cadre d'analyse sémiologique du *cours d'expérience* permet de décrire le mode selon lequel ce vécu s'inscrit dans la dynamique de signification d'un acteur à un instant donné (Theureau, 2006). Cela en se fondant sur l'hypothèse que le contenu de ce qui est exprimable à un instant donné par un acteur déborde la simple description de ce qui advient à cet instant, car il est toujours relié à l'activité précédente et successive. En ce sens, Theureau (2006) souligne la dimension dynamique et « ouverte aux deux bouts » de l'activité, un flux qui hérite à chaque instant de l'histoire de l'activité passée (concept husserlien de « rétention ») et qui s'ouvre sur le futur (concept husserlien de « protention »). Le rapport entre protention et rétention est asymétrique en faveur de la protention : le vivant étant fondamentalement normatif et l'activité étant son expression, elle est d'abord orientée vers le futur immédiat (Husserl, 1964). À la lumière de ces conceptions, nous pouvons considérer l'impasse comme un blocage dans la rétention du passé qui paralyse la protention vers l'avenir.

## 2. Accompagner la relance de la sémiose et étendre l'entendement en formation

Les deux études que nous allons présenter ici permettent de saisir que, dans le vécu d'impasse comme le repli ou « l'effondrement » (Weick, 1993) de la dynamique de signification[5], une issue ou un rebond sont rendus possibles par un mouvement abductif. Dans le premier cas, nous analyserons une formation à visée « réparatrice » visant à aider les participant-es à dépasser un vécu d'impasse et à redéployer leur sémiose. Dans le second cas, nous analyserons une formation à visée « préparatrice », dans laquelle le vécu d'impasse est produit « artificiellement » afin d'aider les participant-es à se préfigurer de façon maitrisée une situation qui pourrait potentiellement se produire à l'avenir en contexte réel, et pour travailler en conséquence à étendre leur entendement.

### 2.1. Un prototype de formation « réparatrice » à médiation artistique

À titre d'exemple d'une formation ayant une visée « réparatrice », nous présentons le dispositif de formation « Théâtre du vécu » (TdV) (Assal, Durand, & Horn, 2016) adressé à des adultes aux biographies marquées par des pathologies graves ou des épisodes dramatiques. Ce dispositif, propose une élaboration par étapes de certains évènements de l'expérience des participant-es, et sa transformation en une forme théâtrale qui lui donne une signification nouvelle. Il se démarque d'une perspective de formation dite curriculaire et centrée sur des objectifs d'apprentissage, et s'inscrit dans une tradition humaniste de l'éducation visant le développement des adultes (Salini & Durand, 2016).

Le TdV consiste en l'accompagnement de cinq ou six participant-es pendant trois jours dans un espace théâtral, par des professionnels de l'éducation des adultes, du soin et du théâtre. Il est demandé aux participant-es : (i) d'écrire un texte concernant « un épisode important de leur vie », (ii) de le mettre en scène, (iii) de diriger les comédiens jusqu'à la réalisation du spectacle devant un public restreint. Des « tables rondes » rythment chaque phase du TdV. Les participant-es y témoignent de l'expérience qu'ils vivent, commentent celle de leurs pairs, et contribuent à une réflexivité partagée. L'atelier se déroule dans une dynamique de soutien et aide de la part des intervenants et de support mutuel et de solidarité entre participant-es.

L'étude qui a concerné ce dispositif repose sur l'observation ethnographique de cinq ateliers de TdV et l'enregistrement vidéo de trois d'entre eux (Salini & Durand, 2016). L'activité de 18 participant-es a été analysée en détail à partir d'observations in situ et de séances d'Autoconfrontation (AC) aux traces de ces ateliers (résumés vidéo, notes de terrain) (Theureau, 2010). Ces séances[6], elles aussi enregistrées, facilitaient l'expression de leur expérience par les acteurs, opérationnalisée comme ce qui en est montrable, racontable, mimable et commentable à tout instant de leur activité (Theureau, 2010). Lors des séances, les acteurs suivaient le déroulement de leur activité passée sur la vidéo, pour exprimer leur expérience à ce moment-là, en évitant les explications, généralisations et justifications. Ces données ont été complétées par un à trois entretiens post TdV, au domicile des acteurs, à des intervalles de trois mois à deux années.

---

5  À considérer « qu'émergence de signification » au sens de Varela et « sensemaking » au sens de Weick soient assimilables, ce qui n'est pas acquis et nécessiterait des travaux comparatifs (Récopé, Fache, Beaujouan, Coutarel, & Rix-Lièvre, 2019).

6  Comme chaque atelier de TdV dure environ 24 heures, on ne pouvait pas soumettre aux participant-es l'ensemble des enregistrements. On a ainsi constitué des extraits d'une heure environ, spécifiques à chacun, en référence aux étapes fondamentales de l'atelier même.

Le traitement a consisté en une analyse sémiologique de l'expérience des acteurs impliqués sur la base du signe hexadique. Les éléments exprimés par chacun ont été d'abord organisés en chroniques individuelles structurées selon la temporalité et les activités spécifiques de la trame du TdV, puis analysés selon le cadre d'analyse sémiologique présenté auparavant (Theureau, 2006). Le traitement final visait à identifier des composantes de l'expérience évaluées comme typiques aux niveaux individuel et inter-individuel. Les jugements de typicité étaient posés à partir des critères de fréquence d'occurrence dans l'échantillon enquêté, et de jugements de typicité émis par les acteurs lors des AC.

Les résultats de cette étude permettent de saisir les multiples éléments qui contribuent au dénouement de certains vécus d'impasse des participant-es. Ils seront présentés en s'appuyant tout particulièrement sur l'expérience de Júlia qui a suivi un de ces ateliers en 2013. Elle a été rencontrée à deux occasions : lors de l'AC dans la semaine suivant l'atelier et six mois plus tard.

### 2.1.1. Je ne suis pas un héros

Júlia[7] est une soignante expérimentée qui, lors de l'atelier, souhaite démêler une situation très complexe concernant les jours où elle se confrontait à la tentative de suicide de la mère d'un enfant dont elle s'occupait, ainsi qu'aux critiques de ses collègues eu égard à son supposé « excès d'implication ». Son vécu d'impasse concerne à la fois ce qui pour elle est « impensable », c'est-à-dire le geste de cette mère et aussi l'impossibilité de sortir du questionnement par rapport à sa façon d'agir et de s'investir de manière importante auprès d'elle. Elle reporte son dilemme dans son texte initial, dont nous reprenons ci-après quelques extraits.

> *« Je ne suis pas un héros*
> *Je sors de sa chambre, je suis bouleversée (…) La première chose qu'elle m'a demandé a été : "et les enfants ? " Je lui parle, lui parle. Son regard est franc et des larmes commencent à couler. Je pleure avec elle et, tout en essuyant ses larmes, je lui dis et redis : "je suis vraiment désolée" (…)*
> *Depuis deux ans nous sommes toute une équipe qui soutenons cette famille. Tout a été fait pour les soins de son enfant. (…) Mais j'essaye de ne pas la juger…*
> *Et là les phrases fatidiques des collègues, que je déteste : "Fais attention, protège-toi, est-ce que tu n'en fais pas trop ?". Qu'est-ce que ça m'énerve ! (…). De même, je pense à moi qui ne peux pas comprendre qu'on puisse ne plus avoir espoir. (…)*
> *Plus tard, je lui dis que si elle n'est pas morte c'est que ce n'était pas le moment, qu'elle a encore des choses à faire… Là on éclate de rire, son rire est un cadeau qu'elle me fait, elle me dit merci car, même si je ne comprends pas son acte, je ne la juge pas (…). »*

### 2.1.2. Dénouement de l'impasse et relance de la sémiose

Nous reportons ici quelques moments-clés de l'expérience de Júlia pendant le TdV, documentés lors de l'entretien d'AC qui a suivi. Nous évoquons aussi les principales composantes du signe hexadique identifiées.

Comme indiqué auparavant, l'Engagement (E) de Júlia concernait la possibilité, lors de l'atelier, de démêler, de mieux comprendre et se faire comprendre, par rapport à des sentiments multiples qui la traversaient lors de la situation difficile choisie. Elle s'implique dans l'atelier (qui lui a été proposé par un médecin de sa connaissance), confiante, mais aussi dubitative sur la manière de s'y prendre. L'ambiance proposée par les intervenants lui fait signe (R) de manière positive et ainsi elle l'interprète (I) : « *C'est un travail intelligent pour nous mettre à l'aise, je suis vraiment très vite à l'aise et très vite mise en sécurité* ». L'accueil (R) de la part des autres participant-es, la touche, car par de petits gestes on fait sentir qu'on est là, présent quoique silencieux : « *un moment qui*

---

[7] Ici et dans l'étude suivante, les prénoms ont été changés.

*m'émeut encore, un des autres a juste mis la main sur mon épaule et il me dit : vas-y, tu as le droit ».*

Comme pour la plupart, pour elle, l'écriture se révèle plus facile que prévu : « Je me disais : comment ça ira ? (U/I) (…). Je suis étonnée parce que je m'attendais à ce que ça soit difficile pour moi (A). Mais non, ça coule de source ». Le soir du premier jour, entendre la lecture des textes des autres est aussi émouvant « ça te sort de ton histoire et ça te transporte dans des situations (I). J'essaie de comprendre ce que chacun vit (E) et je suis très concernée par les autres textes (R) ». De même la lecture neutre par les comédiens ouvre sur une tout autre expérience : « Il y a surtout cette prise de distance. De se dire : il y a quelqu'un d'autre qui raconte (R), c'est comme si on observait la situation, comme si on flottait un peu » (I).

Lors de cette lecture, ainsi que tout au long de la mise en pièce, on note des préoccupations relatives à la qualité et à la fidélité du texte : « Est-ce que je suis assez fidèle à ce qui s'est vraiment passé ? (E). Et j'observe : ça va, ça va ce que j'ai dit (I). Comme une vérification : genre tu n'as pas dit trop de bêtises (I) ». Pendant la mise en scène, il y a aussi de l'inquiétude sur la manière de s'y prendre (I/U). À ce moment, le soutien du metteur en scène est crucial : il accompagne et suggère tout en laissant la liberté de choix : « ses questions : "est-ce que c'est vraiment comme ça ? tu veux une autre couleur ? tamiser la lumière ?" Toutes ces questions font qu'il me donne cette permission de prendre soin de ma mise en scène ». La question de la fidélité revient tout le temps. Par exemple, lorsqu'un comédien dit une chose très dramatique (R) « je me dis : "ça ne me ressemble pas" (U)… je me dis que je tiens à ce que ça soit fidèle à ce que j'ai ressenti de la situation (E) ». La préoccupation esthétique est également très présente, car pour Júlia il faut aussi que la pièce garde une dimension de beauté (E).

Tout au long de l'élaboration, la pièce « dialogue » avec son auteur et propose aussi des possibilités de compréhension et de lecture inédites par les proches « qui comptent » (Salini & Durand, 2016). Et d'un coup Júlia se dit, pendant le visionnement de sa pièce complétée : « *il faut que ma collègue voie ça. (A) Qu'elle puisse voir ce qu'il s'est passé. Et qu'on puisse en discuter toutes les deux* (E) ». Ceci se réalisera en fait, et finalement Júlia ressent d'avoir été comprise par sa collègue : « *Quand elle a pu voir, elle a dit " Merci. Maintenant je comprends. Je n'ai jamais osé te dire mes sentiments contradictoires, toi tu les avais aussi et on n'a jamais partagé ça tout bêtement"* ». De même, pour Júlia l'atelier a eu un effet d'éclaircissement et de confirmation de ses valeurs : « *Le fait de l'avoir posé* (l'épisode évoqué) *ça le met en relation avec mes valeurs et que c'est OK avec moi* (I) ». Par ailleurs, elle se sent mieux dans les rapports avec la patiente évoquée ainsi qu'avec d'autres également complexes : « *Quand je les vois, il y a plus de sérénité dans mon attitude* (I). *Je ne suis plus phagocytée par ces pensées. Je ressens une espèce de calme, de sérénité* (E) ». La rencontre avec soi-même, par un effet de connaissance de soi et d'ouverture d'un potentiel valant pour le futur, est aussi soulignée : « *je me connais mieux et je vois que je suis comme les autres* (I) ».

Grace au TdV la difficulté initiale et le vécu de souffrance sont « re-présentés » en tant qu'évènement marquant dans le texte-récit rédigé le premier jour du TdV. Il s'agit de textes simples, qui vont à l'essentiel, mais leur rédaction permet de donner une organisation et une consistance à ce vécu en adoptant un schéma narratif, tout en amorçant son extériorisation.

L'élaboration de la pièce est une occasion de « revenir en arrière » sur la situation à l'origine de sa difficulté, tout en ouvrant vers un avenir, grâce au travail de métaphorisation engagé. L'ensemble de la mise en scène évoque de manière iconique les différentes facettes du vécu de Júlia et son possible dénouement. Par la voix des acteurs elle raconte ses doutes, ses contradictions, ses angoisses et questionnements. Lors de l'élaboration de la pièce, les émotions vécues reviennent comme « saisies de l'extérieur ».

Júlia fait alors simultanément l'expérience d'être auteur et spectateur de son œuvre. Dans ce jeu d'engagements multiples, une expérience transitionnelle complexe se produit : d'une préoccupation à une autre, d'une émotion à une autre, d'une signification à une autre. Les moments d'échange qui suivent chaque moment-clé de l'atelier, ces « tables rondes », permettent à Júlia de partager ses préoccupations. Grâce à la pièce, les autres participant-es ont ressenti son vécu, mais ont également saisi ses valeurs essentielles. Son expérience individuelle se transforme par ce travail en présence des autres ainsi que par l'écoute et l'accompagnement de tous.

Plus en général, dans le TdV, la médiation de l'expérience-cible (le vécu accablant) est rendue possible par l'amplification de l'activité imaginative et par l'enchâssement de plusieurs éléments de nature fictionnelle (Salini & Durand, 2016). Ces éléments peuvent soutenir la conception de dispositifs de formation à médiation artistique, et nous les synthétisons ainsi :

— Le registre *mimétique*. Les participant-es, grâce à la production de la pièce, entrent en résonance à la fois avec leurs vécus et ceux des autres (Gallese, 2011) ;
— Le registre du *musardage* (Peirce, 1994). La proposition théâtrale permet aux participant-es de se libérer d'une finalisation étroite de l'action, et de faire place aux errements : on fait confiance à l'imagination où l'amusement et le rire ne sont ni scandaleux ni temps perdu ;
— Le registre de la *métaphorisation*. Le texte produit initialement, qui reste toujours la référence pour l'élaboration scénique, devient tridimensionnel et s'enrichit de possibilités d'expression inédites tout au long de l'atelier grâce à l'omniprésence de métaphores. Celles-ci sont conventionnelles et culturelles. Elles facilitent la compréhension par le spectateur de ce dont il est question parce que la mise en scène « dit » à chaque instant : « ceci est une mise en scène de cela ». Ces métaphores ouvrent au possible dénouement de l'événement (au sens propre de défaire un nœud) qui constitue l'objet de la mise en scène. La constitution progressive de la pièce théâtrale ouvre ainsi la possibilité de voir des situations qui sont restées « en suspens » trouver leur épilogue, ou de rendre de façon imaginaire des absences et des vides d'échanges qui ont été vécus ;
— Le registre d'implication *métaleptique* (Genette, 2004). Comme il ressort de l'expérience de Júlia, chaque participant-e exprime divers engagements simultanés ou successifs au regard de son texte et de sa production théâtrale : (i) comme témoin en première personne et interprète de son expérience, (ii) comme auteur d'un texte le concernant, (iii) comme metteur en scène de ce texte, (iv) comme spectateur-commentateur et analyste de son propre spectacle, (v) comme spectateur-commentateur et analyste du spectacle des autres, et enfin (vi) comme apprenti du processus de mise en scène et de direction d'acteurs. Ces postures différentes sont sources d'intrusions et de décalages pendant l'atelier, suscitant des ruptures et des télescopages qui favorisent des rapports changeants à l'œuvre en cours et aux expériences personnelles ;
— *L'externalisation du dialogue interne* (Rosenthal, 2012). La dynamique de signification est « intrinsèquement sociale au sens où elle s'adresse à un interlocuteur au moins virtuel » (Chauviré, 2016, p. 118), la dimension théâtrale du TdV, notamment par la création de personnages et d'alter ego, concrétise et extériorise le dialogue interne (Peirce, 1994) ainsi que les doutes, les contradictions ou les hypothèses qui les traversent. Ceci permet aux participant-es de les voir du dehors en ayant ainsi la possibilité de les ressaisir à nouveaux frais.

## 2.2. Un prototype de formation « préparatrice » : l'exercice de crise

### « hors cadre »

Envisagé sous l'angle formatif, le second dispositif étudié est un exercice de crise « hors cadre », c'est-à-dire simulant un événement totalement inédit, et analysé comme prototype de formation ayant une visée « préparatrice ». Elle vise à aider les individus impliqués à développer des dispositions « à voir », « à percevoir », « à sentir », « à agir », « à imaginer », « à interagir » dans des situations futures, qui, sans cela, sont susceptibles d'être vécues comme critiques et d'amoindrir voire d'annihiler leur capacité d'attention, d'interprétation et d'action (stress, incertitude, effondrement de la sémiose…), en particulier dans les contextes à risques (ici la sécurité industrielle). L'exercice « hors cadre » consiste à scénariser des événements « inattendus », mais également « impensés » et vise à faire expérimenter aux individus les situations engendrées et le vécu d'impasse associé par ces types d'événements, de façon maitrisée.

Cette étude est issue du projet FOResilience[8] portant sur la formation *à et par la résilience* dans le domaine de la sécurité industrielle. Notre analyse porte d'abord sur le comportement d'un acteur-clé primo-intervenant (l'agent de conduite en responsabilité, installé au pupitre de supervision) au cours d'un exercice de crise pleine échelle organisé sur un site de stockage de gaz. Les données de cette étude ont été construites à partir d'une observation ethnographique et d'un entretien consécutif avec cet acteur suivant la méthode de verbalisation décalée (Theureau, 2010). À l'aide du signe hexadique nous rendons compte du dépassement d'un vécu d'impasse par abduction, puis nous dégageons les opportunités de développement favorisées par la confrontation à un événement se révélant non seulement « inattendu », mais également « impensé ».

### 2.2.1. « Vous ne me faites jamais ça ! »

Le cas présenté ici est particulièrement intéressant pour l'un des événements de son scénario, conçu pour dérouter les opérateurs : la simulation de la mort de deux agents d'astreinte. Au cours de l'exercice, l'événement se produit comme suit : alerté d'une anomalie sur une conduite de gaz (R), l'agent de conduite applique la procédure (S) en contactant les agents d'astreinte pour leur demander d'effectuer une « levée de doute » sur place (U/S). Quelques minutes plus tard, ces agents décèdent fictivement dans l'explosion de la conduite, mais comme prévu par le scénario aucun participant-e n'est présent sur place pour en rendre compte. L'animateur de l'exercice place les deux agents d'astreinte hors exercice, et informe seulement les autres opérateurs, dont l'agent de conduite, qu'une explosion de forte intensité a retenti dans la zone (R).

L'activité de l'agent de conduite, primo-intervenant et à cet instant principal acteur de la gestion de crise, est alors perturbée dans le domaine de l'« inattendu », car les explosions sont très rares (S), mais tout à fait possibles (S). Il applique alors la procédure prévue (S) en contactant les pompiers (U). La perturbation bascule dans le domaine de l'« impensé » lorsqu'il cherche à joindre par téléphone les deux agents d'astreinte (E). Après avoir essayé quatre fois en vain (U/R) – alors que l'astreinte a l'obligation formelle de répondre aux appels (S) – il est envahi par un stress manifeste et une incompréhension (I) qu'il traduit verbalement (U) : « C'est pas vrai ?!! » (I) (7h01 et 7h05)[9], « Mais rappelle-moi ! » (U) (7h06), « C'est pas possible ! » (I) (7h07), « On va pas y arriver… » (I) (7h09).

Les hypothèses d'une défaillance technique du système de communication (I) ou d'un écart de procédure de la part de l'astreinte (I) lui semblent tout à fait improbables (I). Il émet alors une hypothèse qui lui semble plausible (I) : une faille dans le scénario de l'exercice. En effet, il estime qu'il n'est pas *possible* que l'astreinte ne réponde pas au téléphone (I), cela même s'il a été informé de l'explosion (S). Mais l'animateur lui confirme le bon déroulement du scénario (R), ce qui le maintient dans un état de stress et de perplexité (R/I).

---

[8] Pour plus d'informations concernant le projet FOResilience :
https://www.unige.ch/fapse/craft/recherche/passees/foresilience/

[9] Ces indications horaires sont les moments de l'exercice auxquels ces verbalisations ont été produites.

Il se met alors à chercher comment poursuivre son action dans cette situation inédite (E) bien que cela implique de s'écarter de la procédure (I) (levée de doute par l'astreinte), et d'accepter au moins momentanément son état d'incertitude (I). Il renonce à contacter l'astreinte (U) et décide de contacter un autre agent de maintenance (U) qui se trouverait à proximité de la zone sinistrée. Il consulte le planning dynamique des opérations (U) qui indique avec une grande précision et une bonne lisibilité où sont et que font les opérateurs à chaque instant (S). Il identifie un agent (R) et le contacte pour lui demander de se rendre sur place (U). Un aspect sécuritaire émerge alors dans son champ de préoccupation (E), et il conclut sa demande en disant avec insistance (U) : « *Et tu fais gaffe, hein ! Ça a explosé là-bas… Tout d'un coup j'ai peur pour Cédric et Thomas* » (I) (7h11). Il prend alors conscience que les agents d'astreinte peuvent avoir été blessés dans cette explosion au point de ne pouvoir répondre au téléphone (R/I). Il recontacte alors les pompiers (U) en précisant qu'il craint désormais une urgence vitale (I).

Après la phase simulée, au début du débriefing, l'agent de conduite lance aux deux agents d'astreinte « *Vous me faites jamais ça, hein ?* » (I), ce qui montre le réel retentissement qu'a eu pour lui l'expérience vécue en contexte fictif. Les opérateurs de la cellule de crise relatent également d'emblée que simuler l'étape d'appel téléphonique des conjointes des agents fictivement décédés les a profondément déconcertés.

**2.2.2. Dépassement de l'impasse et extension de l'entendement**

Nous analysons ici le comportement de l'agent de conduite vis-à-vis de la perturbation induite par l'événement « hors entendement » qu'il a vécu (impossibilité de joindre l'astreinte) afin de dériver les opportunités de développement favorisées par les domaines de scénarisation de « l'inattendu » puis de « l'impensé ».

Lorsqu'il cherche, à quatre reprises, à joindre par téléphone les deux agents d'astreinte (en vain), l'agent de conduite manifeste un haut degré de stress et de confusion. Cet épisode est une bonne illustration de la nécessaire distinction à opérer entre *disposer des savoirs nécessaires à l'interprétation et à l'action efficace*, et *produire des significations assurant une intelligibilité des événements*. En effet, l'agent de conduite sait évidemment (i) qu'il y a eu une explosion, et (ii) que les agents de terrain sont vulnérables aux explosions, donc il dispose de tous les savoirs nécessaires ; néanmoins il s'avère incapable à ce moment de produire une interprétation vraisemblable, ce qui le paralyse.

Cette perturbation conduit donc successivement l'agent de conduite : (i) à un état initial de désarroi,« d'effondrement » de la sémiose et de paralysie ; (ii) à accepter son état d'incertitude et à chercher de nouvelles ressources et de nouveaux moyens pour atteindre son objectif, même si cela implique de s'éloigner de la procédure ; et (iii) à faire par abduction plusieurs hypothèses invalidées pour l'action avant de parvenir à une hypothèse permettant provisoirement de sortir de l'impasse (et qui se révélera être en phase avec le scénario).

À partir d'inférences théoriques, nous pouvons penser que cet épisode contient les opportunités d'apprentissage-développement suivantes :

*Domaine de l'« inattendu » (entrée des événements dans la structure d'attentes de l'agent) :*

— Préfiguration d'un cas de figure plausible (e.g., une explosion), dont on peut penser qu'il créera moins la surprise chez l'agent en situation d'accident réel.
— Bénéfice d'un ensemble d'expériences de perception et de significations (émergence

d'émotions, d'indices, de repères, de ressources) liées à ce cas de figure, dont on peut penser qu'elles accroîtront la capacité de l'agent à produire des significations en situation d'accident réel, et ce faisant, son degré de réactivité.

— Préfiguration d'une solution soutenable (e.g., levée de doute par un agent tiers) qui, ainsi *préfigurée,* sera plus susceptible de se *reconfigurer* de manière semblable en situation d'accident réel.

*Domaine de l' « impensé » (entrée des événements dans la culture propre de l'agent) :*

— Extension de l'entendement, c'est-à-dire de la capacité à produire des significations (i) *sur* le critique (ce qui peut arriver, e.g., la mort d'un collègue) et (ii) *dans* le critique (ce qu'on peut faire, e.g., s'écarter de la procédure et improviser un mode opératoire inédit).

— Développement d'une sensibilité à une « possibilité de l'impossible » et à une nécessité d'improvisation : si la mort d'un collègue n'est pas impossible, d'autres évènements supposément impossibles ne le sont peut-être pas non plus.

— Identification d'une « bouée de sauvetage », d'une ressource toujours disponible à laquelle se raccrocher en cas d'urgence et de confusion (e.g., le planning dynamique des opérations).

Ces opportunités de développement sont favorisées par des principes de conception comparables à ceux du Théâtre du vécu, tout en s'actualisant différemment.

## 3. Principes de conception de formation

Malgré leurs nombreuses différences de prime abord (contexte, public, objectifs, modalités...), les deux dispositifs présentés peuvent être saisis au prisme de principes de conception communs qui sont de nature à provoquer, favoriser, soutenir la construction de nouvelles significations face à des évènements inédits et critiques. Nous regroupons ces principes sous deux thèmes principaux : celui de la dimension fictionnelle et celui de la dimension événementielle.

Bien qu'elles soient entrelacées, on peut distinguer deux niveaux d'inscription de ces dimensions dans l'hypothèse d'activité-signe. La dimension fictionnelle se rapporte notamment aux dimensions iconiques de la signification (en référence à la composante Interprétant du signe hexadique), comme les métaphores présentes dans les deux études mobilisées (dont la scénarisation est l'élément transversal). La dimension événementielle se rapporte notamment à la surprise, à l'inattendu et à l'impensé comme des facteurs d'enquête dont l'aboutissement espéré est la construction de nouvelles significations.

### 3.1. Dimension fictionnelle

Bien que fictives, les situations de gestion de crise (simulées par le scénario) et les situations difficiles ou traumatiques (transformées en fiction par les comédiens) donnent lieu à des expériences bien réelles qui marquent les individus concernés, potentiellement durablement, en leur faisant découvrir, interpréter et agir dans des situations encore « impensées » et « inéprouvées ».

Cette dimension fictionnelle de l'expérience rendue possible par des simulations de nature différente permet de faire advenir des situations dont le caractère de rareté (épisodes dramatiques actualisés dans le TdV, épisodes de crise anticipés dans l'exercice) rend l'appréhension incertaine et difficile. Mais les faire advenir dans le domaine de l'expérience des participant-es de façon identique à ce qu'elles ont été ou pourraient être ne serait non seulement pas formateur, mais potentiellement délétère. Le potentiel de formation réside dans une médiation de l'expérience-cible (artistique en ce qui concerne le TdV, ludique[10] pour l'exercice de crise) qui lui confère à la fois un caractère de familiarité (nécessaire à l'obtention d'un effet mimétique entre formation et vie quotidienne), d'étrangeté (nécessaire à l'obtention d'une enquête de la part des participant-es) et de sécurité (nécessaire à un engagement authentique de la part des participant-es). Cela permet aux participant-es d'expérimenter de façon relativement maitrisée « ce que ça peut être », « ce que ça peut faire », « ce que ça peut vouloir dire » (d'être confronté à nouveau dans des circonstances différentes à un événement critique vécu – dans le TdV –, ou bien d'être confronté pour la première fois à un événement critique possible – en exercice). Ces conditions sont très favorables à l'émergence d'un processus abductif.

L'activité des participant-es dans le TdV et dans l'exercice de crise partage une caractéristique que Horcik, Savoldelli, Poizat et Durand (2014) définissent comme une « double intentionnalité ». En effet, la dimension fictionnelle à l'œuvre dans ces deux dispositifs génère ou révèle chez les participant-es des engagements enchâssés, et orientés – soit simultanément soit alternativement – vers l'enjeu de la formation (faire aboutir l'enquête scénarisée) et vers l'enjeu ciblé dans la vie ordinaire (dépasser une impasse). Le TdV ressemble à du théâtre (écriture et jeu d'une pièce), mais sans en être (la finalité et l'adressage sont très différents du théâtre ordinaire). L'exercice de crise ressemble à une gestion de crise (événements, procédures, stress…), mais sans en être une (on sait toujours qu'on est en simulation et que les erreurs n'auront pas de conséquences réelles pour la sécurité). Ces auteur-e-s conceptualisent l'expérience afférente à cette « double intentionnalité » comme une « double négation » : le TdV n'est *pas* du théâtre, mais n'est *pas pas* du théâtre ; l'exercice de crise n'est *pas* une gestion de crise, mais n'est *pas pas* une gestion de crise. Le vécu qui accompagne cette tonalité d'engagement fictionnel est estimé particulièrement prometteur pour favoriser des pontages entre formation et vie ordinaire (Durand, 2008).

### 3.2. Dimension événementielle

Un évènement est une occurrence qui rompt le cours de la vie ordinaire. Il ne s'agit pas simplement d'une circonstance qui prend sens « en soi », mais d'un fait saisi comme imprévisible dans le flux du quotidien. C'est une discontinuité qui, tout en étant immanente à une situation, suspend le temps et marque un arrêt entre passé et avenir, ouvrant une brèche dans les significations associées à l'existence vécue. Ce fait devient évènement au moment où (i) il s'avère très significatif pour un certain individu – ou une certaine collectivité – qui se sent concerné par lui, (ii) il ne se laisse pas comprendre à partir des significations préalables, mais ouvre à d'autres significations possibles, (iii) il n'est pas saisissable selon un ordre linéaire de cause-effet (Romano, 1998), et (iv) dans le cas où il s'accompagne d'une déstabilisation majeure chez l'individu : il tend à générer un processus d'enquête visant à recouvrer une nouvelle stabilité.

Deux formes possibles d'émergence d'évènements sont suggérées par Zarifian (1995) : celui ayant un caractère aléatoire et celui qui se présente comme *rendez-vous* dans un réseau d'acteurs. Le premier, l'évènement *aléa*, est subi, il fait partie des faits « inattendus » auxquels il faut faire face, et il est souvent associé à des incertitudes, voire des crises. Cet événement demande à être absorbé dans l'existence par une nouvelle configuration des significations. Si ce n'est pas le cas, un questionnement « qui tourne en rond » à son propos s'installe et cet événement représente alors une gêne (plus ou moins importante), quelque chose comme un caillou dans la chaussure, qui fait que des individus s'y arrêtent et cherchent à lui conférer une signification singulière.

L'événement *rendez-vous* relève de l'intention et du souhait d'un accomplissement futur par un réseau d'acteurs qui sont, dans les deux cas décrits plus haut, les formateurs et formatrices engagé-e-s dans les deux dispositifs de formation. Quoiqu'anticipé, l'accomplissement de cet événement est toujours plus ou moins incertain, et sa prédiction totale et minutieuse est impossible. Le vécu de l'événement *rendez-vous* est celui de la projection et de l'anticipation et il vaut comme un engagement à faire quelque chose. Il possède un pouvoir de transformation des significations de l'histoire dont la personne (dans ces cas, la ou le participant-e) est l'héroïne. Cet événement-là possède un pouvoir de génération de changements des dynamiques de signification.

---

[10] La médiation est ludique au sens où en exercice (et en simulation en général), les participant-es *jouent* en faisant « *comme si* » ; elle n'est pas ludique au sens où l'exercice serait amusant.

Dans le TdV, une circularité émerge entre événements *aléa* et *rendez-vous* (Zarifian, 1995). En choisissant ce sur quoi porte leur œuvre théâtrale, les participant-es désignent quelque chose de marquant, source de souffrance enkystée et d'impasse incompréhensible. Est constitué comme un événement *aléa* ce qui est extrait de l'expérience passée et choisie pour référence de la fiction théâtrale. Sa théâtralisation accompagnée par les intervenant-e-s est une promesse d'événement *rendez-vous* qui peut faire émerger des significations nouvelles, « réparatrices ». On cherche à ce que les situations jouées en formation aient pour les participant-es un « air de famille » avec les situations critiques, la formation contribuant ainsi à une double circulation d'expériences : entre les événements, chronologiquement distants, et entre les personnes présentes, qui ne sont pas les mêmes dans la situation passée et présente.

Dans l'exercice de crise, le mouvement est inversé : on vise une circularité entre un événement *rendez-vous* (simulation en formation) antérieur à l'*aléa* (risque que survienne une crise). Ceci advient en proposant des situations qui perturbent le couplage acteur/environnement par la provocation d'une expérience critique maitrisée. Pour ce faire il faut d'une part doser le caractère critique de la perturbation, de l'autre offrir des possibles par proscription, selon le principe : « tout ce qui n'est pas interdit est autorisé ». D'ailleurs, le dépassement de la situation critique, elle-même devenue évènement *aléa* (pour les participant-es à l'exercice) est rendu possible par l'accompagnement à un « dégagement analytique » post-événement. Ceci dépasse le traditionnel débriefing (écart entre attendu et réalisé), peut débloquer un état de désorientation, de renoncement ou de rupture, et favoriser une forme d'activité soutenable et une capacité de développement et de projection vers l'avenir. On cherche ainsi à ce qu'une hypothétique future situation de crise ait pour les participant-es un « air de famille » avec une ou des situations vécues directement – ou par procuration – par le biais de formations et qui ont fait émerger des significations nouvelles, « préparatrices ». Ceci contribue ainsi également à une double circulation d'expériences (entre les événements et les personnes), mais dans le sens opposé.

Dans les deux cas, la dimension événementielle s'avérant très favorable aux processus abductifs, les significations produites initialement sont *reenactées* dans des conditions différentes. Ainsi, elles sont réélaborées selon des modes de renforcement, d'affaiblissement et/ou de relativisation qui tendent à enrichir le Référentiel des participant-es, envisagés comme des « réservoirs de sens » par Weick (1993), et ainsi leur capacité à signifier. Cette sémiose tend à aider les participant-es à « réparer » un maillon d'une chaine interprétative (dans le TdV), ce qui permet de mieux vivre avec un épisode traumatique, et/ou à créer de nouveaux maillons (dans le TdV et l'exercice), ce qui permet de recouvrer/développer une capacité de projection vers l'avenir.

La dimension événementielle exploitée par les deux dispositifs se déploie également au niveau social à travers les communautés qui participent, que ces communautés leur préexistent (communauté de pratique et de métier en exercice de crise) ou non (génération d'une communauté par un engagement fort et commun dans le TdV). Dans les deux cas, l'un des rôles de la scénarisation est de faire collectivement événement, c'est-à-dire de dramatiser les circonstances pour faire naître chez les différent-e-s participant-es un intime sentiment de « concernement » (Brunet, 2008). Celui-ci consiste en un engagement intentionnel et affectif « de fond » envers l'enquête et ceux qui la partagent, qui se traduit notamment par une attention soutenue à soi et aux autres. Plus que dans de nombreux autres dispositifs de formation, la nécessité que le collectif réussisse est primordiale et passe par le fait que chacun-e « s'en sorte ». Dans le TdV, l'événement collectif émerge de dynamiques collectives riches, soudées au plan émotionnel et notamment soutenues par la mutualisation de vécus individuels très chargés affectivement. La mise en scène exploite ces dynamiques pour créer les conditions d'une expression esthétique et d'une présence attentive à l'autre. Ces phénomènes ne sont pas seulement les prémisses de la relance de la sémiose au plan individuel : ils en sont le véhicule.

Dans l'exercice de crise que nous avons analysé, l'événement de la mort fictive de leurs collègues de terrain génère chez les agents une double expérience significative de vulnérabilité au risque industriel : celle de leur propre vulnérabilité à l' « impensé » (en éprouvant de la confusion, un sentiment d'impuissance dans une situation qui échappe à leur compréhension voire à leur entendement) ; mais aussi celle de la vulnérabilité physique des autres, collègues, amis (en faisant l'expérience fictionnelle d'une situation possible : l'accident mortel). Ces phénomènes ont une fonction développementale contre-intuitive, parce qu'on pourrait penser que l'accroissement du sentiment de vulnérabilité est une régression. Néanmoins, ils participent au développement d'une humilité et d'une vigilance en actes, dispositions que les sciences de la sécurité identifient comme très favorables à la construction continue des conditions de la sécurité (Weick & Sutcliffe, 2001).

## 4. Contribution au programme de recherche technologique en formation et au programme du CdA

Nous avons essayé d'expliquer dans cet article que, pensée sous l'hypothèse d'activité-signe, la formation peut dans certains cas (notamment ceux d'impasse, que nous avons décrits) avantageusement spécifier ses objets et ses méthodes à partir d'une compréhension de différents modes de perturbation et/ou relance de la sémiose et de leur articulation. Nous en discutons ici les implications d'une approche par la perturbation de l'activité en formation à des fins de reconfiguration des significations.

Nous avons décrit, à partir des deux formations analysées, les opportunités de développement suscitées par la prise en compte ou l'utilisation de perturbations dans le vécu des acteurs, face à des évènements inédits ou critiques. Les dimensions « réparatrices » et « préparatrices » concernent la perspective pour les individus concernés soit de re-signifier, soit d'avoir à gérer des situations critiques relevant des domaines de l'« inattendu » et/ou de l'« impensé ». Elles sont contenues dans deux composantes communes (et les effets associés) aux formations étudiées, tout en se déployant de manière différente pour l'une et l'autre.

La première a déjà largement été décrite comme un effet d'extension de l'entendement, ou « domaine cognitif expérientiel » (Varela, 1996), c'est-à-dire une augmentation par abduction des significations possibles dans le monde propre de l'acteur. Ceci est de nature (i) à permettre le dénouement de l'impasse face à des vécus pétrifiants et (ii) à réduire le risque de surprise et a fortiori de sidération ou « d'effondrement » de la sémiose dans des situations n'offrant pas de possibilité de mobiliser et mettre en relation des connaissances pertinentes et disponibles.

La seconde n'a pas encore été décrite et concerne une spécificité des dispositions actualisées face à des situations confuses, que l'on peut qualifier de « capacité négative ». Cette notion, empruntée par le psychanalyste Bion au poète anglais John Keats, concerne la capacité « de demeurer au sein des incertitudes, des mystères, des doutes sans s'acharner à chercher le fait et la raison » (Keats, 1993, p. 76). Dans la relation à l'autre (mais plus généralement dans la relation à l'environnement), c'est une disposition à « tolérer la turbulence émotionnelle de *ne pas savoir* ; s'abstenir d'imposer des solutions fausses, omnipotentes ou prématurées à un problème, une situation ambiguë ou une perturbation émotionnelle » (notre traduction, Bion, 1991, p. 207).

> « Essentiellement, il s'agit de faire face à l'expérience plutôt que de l'éviter. [...] En tant que telle , [la capacité négative] décrit la nature de la recherche d'une mentalité […], pensante, qui va au-delà des principes didactiques connus et élargit la structure de la personnalité » (notre traduction, Williams, 2018, p. 42).

Cette capacité est dite *négative* au sens où, lorsqu'un individu parvient à tolérer l'insuffisance perçue de ses moyens d'action et de ses habitudes de pensée pour qu'une situation perturbante soit vécue de manière acceptable, il ouvre une capacité « en creux » (Blanchard-Laville, 2013), une disponibilité à inventer autre chose, c'est-à-dire à l'abduction. Mais cela suppose pour toutes les personnes engagées dans les formations, formateurs-trices et participant-es, (i) d'accepter que leurs compétences, leur motivation, leur sincérité ne garantissent aucunement la réussite du TdV ou la gestion efficace d'une crise « hors cadre », et (ii) d'être attentives et attentionnées envers elles-mêmes, les autres, et ce qu'elles ont à faire ensemble.

Par exemple, les professionnel-le-s de l'éducation, de la guérison et du théâtre qui interviennent dans le TdV sont hautement compétents. Ils assurent aux participant-es du respect et aucun jugement de valeur sur leurs productions ou leur expérience vécue. Ils sont d'ailleurs un peu éloignés de leurs pratiques habituelles : ils font leur travail sans vraiment le faire. Ils sont les garants d'un processus de théâtralisation et accompagnent les participant-es dans la réalisation de leurs œuvres, bien que la plupart du temps ces derniers ne connaissent pas les codes du théâtre. En accompagnant les participant-es, ils doivent respecter leurs goûts, leurs choix artistiques et, surtout, l'expression de ce qui se trouve au plus profond d'eux. Malgré l'incertitude de la situation, ils agissent selon une « capacité négative », ils prennent soin des participant-es, confiants dans leur potentiel de développement et l'impact du processus artistique.

Dans le domaine de la sécurité industrielle, le concept de « capacité négative », qui reste largement à étayer empiriquement et conceptuellement, renvoie conjointement à ceux de « mindfulness » (pleine conscience, vigilance), de « situation awareness » (conscience / compréhension de la situation) et de « sensemaking » notamment théorisés dans les sciences des organisations à la suite de Weick. L'intérêt de ce concept est d'essentialiser et de développer la capacité à agir dans l'inconnu sous un postulat tacite « d'inconnaissabilité » (Weick, 2006), c'est-à-dire que l'opérateur conçoive intuitivement que la compréhension d'une situation par déduction de ses connaissances antérieures ne doit pas toujours précéder l'action : en cas de grande confusion, agir sur la base de repères et de procédés minimaux peut constituer une amorce de compréhension de la situation par abduction (émission et test d'hypothèses provisoires). Déterminante pour penser l'intervention vis-à-vis de situations fortement indéterminées, voire critiques (Pettersen, 2013), cette capacité nous parait également devoir être pensée pour la formation.

Les effets décrits de l'extension de l'entendement et du développement d'une capacité négative favorisent l'expression et le développement de dimensions cruciales de la gestion de situations de crise telles que la capacité d'interprétation (dépasser la sidération, l'incompréhension, et accepter d'agir dans une forte incertitude), d'imagination (émettre et hiérarchiser des hypothèses explicatives sur la base d'informations rares, imprécises, voire potentiellement fausses), d'invention (inventer des modalités d'action nouvelles), et d'improvisation (trouver des ressources de différentes natures dans un environnement qui semble en être dépourvu afin de pouvoir agir). Autrement dit, il s'agit pour un individu de développer une disposition (i) à accepter inconditionnellement sa propre vulnérabilité à l'incertitude, alors que ses habitudes de pensée et d'action le disposent à l'inverse, et (ii) à trouver des moyens d'agir, même sans but précis, mais en créant de fait des opportunités de signification (repérage d'indices), d'élaboration d'hypothèses (abduction) et d'expérimentation de ces hypothèses (enquête).

Atteindre de tels objectifs nécessite de créer les conditions favorables au déploiement d'un régime d'activité à forte composante imaginaire, sollicitant les dimensions abductives et iconiques de la signification qui sont très prometteuses de développement. À cela s'ajoute la préoccupation de dramatiser suffisamment l'enjeu de la formation pour faire événement chez les participant-es, en garantissant toutefois le sentiment de sécurité psychologique nécessaire à un engagement authentique de leur part. Cette préoccupation est par ailleurs primordiale pour toute proposition qui vise à faire advenir l'inédit et le critique dans l'expérience des participant-es de façon dramatique (indirecte, métaphorique, ludique, simulée, mis en scène, narrée, jouée...).

## 5. Conclusion

À partir d'analyses de vécus d'impasse en formation, cette contribution permet de soutenir des choix de conception en matière de formation sur la base de l'hypothèse de l'activité-signe et du processus d'abduction. Elle encourage (i) à ne pas prendre pour objet de formation les seuls apprentissages-développements qui consistent en la construction de savoirs nouveaux de la part de l'acteur, ce qui est réducteur au regard de la diversité des processus d'apprentissage-développement qu'il convient de pouvoir provoquer, capter, ou soutenir dans certains cas, et (ii) à se décentrer chaque fois que c'est opportun d'une logique d'accumulation de savoirs au profit d'une logique de perturbation de l'activité à des fins de relance ou d'extension de la sémiose.

Au final, l'analyse conjointe de ces deux dispositifs de formation particuliers met en lumière l'intérêt des inférences abductives en situation de formation afin de « réparer » et « préparer » à un vécu d'impasse, mais aussi les principes de conception qui encouragent ce processus abductif. Des études complémentaires seraient nécessaires (i) afin de rendre compte et de préciser les processus de construction de nouveaux savoirs par inférences abductives, mais aussi dans l'articulation aux inférences déductives et inductives, et (ii) dans la perspective de tester la fécondité empirique des différentes sous-catégories de l'Interprétant proposées dans le cadre sémiologique du CdA (Theureau, 2015). Ces sous-catégories de l'Interprétant restent à ce jour spéculatives et sont à mettre à l'épreuve dans le cadre de recherche empirique, et tout particulièrement dans les recherches en éducation et formation. Enfin, de futures recherches devront préciser les conditions d'émergence, de génération et d'accompagnement d'abductions majorantes en situations de formation ou de travail. De tels progrès seraient de nature à contribuer à une technologie de formation développée sous l'hypothèse d'activité-signe.

Ahmed, S., & Parsons, D. (2013). Abductive science inquiry using mobile devices in the classroom. *Computers & Education*, *63*, 62-72. https://doi.org/10.1016/j.compedu.2012.11.017

Assal, J-P., Durand, M., & Horn, O. (2016). *Le Théâtre du Vécu : Art, Soin, Éducation*. Dijon: Raison et passions.

Bion, W. (1991). *A Memoir of the Future* (3 Vols. 1975, 1977, 1979). London: Karnac.

Blanchard-Laville, C. (2013). Accompagnement clinique et capacité négative. *Cahiers de Psychologie Clinique*, *41*(2), 63-80.

Bopry, J. (2007). The give and take between semiotics and second-order cybernetics. *Semiotica*, *164*, 31-51.

Brunet, P. (2008). De l'usage raisonné de la notion de « concernement » : mobilisations locales à propos de l'industrie nucléaire. *Natures sciences sociétés*, *16*(4), 317-325.

Chauviré, C. (2016). Le « nous » de Peirce ou la critique de l'égoïsme. *Klēsis*, *34*, 116-129.

Colapietro, V., Midtgarden, T., & Strand, T. (2005). Introduction: Peirce and education: The conflicting processes of learning and discovery. *Studies in Philosophy and Education*, *24*(3-4), 167-177.

Cunningham, D.J. (1998). Cognition as semiosis: The role of inference. *Theory and Psychology, 8*, 827-840.

Cunningham, D. J., Schreiber, J. B., & Moss, C. M. (2005). Belief, doubt and reason : CS Peirce on education. *Educational Philosophy and Theory*, *37*(2), 177-189.

Dieumegard, G. (2009). Connaissances et cours d'expérience vers une grammaire minimale de description dans les situations d'éducation et de formation. *Revue d'anthropologie des connaissances*, *3*(2), 295-315.

Dieumegard, G. (2011). Dimensions cognitives et sociales dans l'étude de l'activité des élèves. La représentation comme inférence individuelle-sociale dans le cours d'expérience. *Éducation et didactique*, *5*(5.3), 33-60.

Durand, M. (2008). Un programme de recherche technologique en formation des adultes. *Éducation & Didactique*, *2*(3), 97-121.

Durand, M. (2017). L'activité en transformation. In J.-M. Barbier, & M. Durand (Eds.), *Encyclopédie d'analyse des activités* (pp. 47-70). Paris: PUF.

Durand, M., Goudeaux, A., Horcik, Z., Salini, D., Danielian, J., & Frobert, L. (2013). Expérience, mimesis et apprentissage. In L. Albarello, J.-M. Barbier, E. Bourgeois, & M. Durand (Eds.), *Expérience, activité, apprentissage* (pp. 39-64). Paris: PUF.

Fisette, J. (2009). L'icône, l'hypoicône et la métaphore : L'avancée dans l'hypoicône jusqu'à la frontière du non-conceptualisable. *Visual Culture*, *14*, 7-46.

Flandin, S., Poizat, G., & Perinet, R. (2019). Contribuer à l'amélioration de la sécurité industrielle « par le facteur humain » : Un regard pour aider à (re)penser la formation. *Collection Regards sur la sécurité industrielle*. Toulouse: FONCSI.

Gallese, V. (2011). La simulation incarnée et son rôle dans l'intersubjectivité. In P. Attigui & A. Cukier (Eds.), *Les paradoxes de l'empathie* (pp. 49‑72). Paris: CNRS.

Genette, G. (2004). *Métalepse : De la figure à la fiction*. Paris: Seuil.

Grossetti, M. (2004). *Sociologie de l'imprévisible*. Paris: PUF.

Horcik, Z., Savoldelli, G., Poizat, G., & Durand, M. (2014). A phenomenological approach to novice nurse anesthetists' experience during simulation-based training sessions. *Simulation in Healthcare*, *9*(2), 94-101.

Husserl, E. (1964). *Leçons pour une phénoménologie de la conscience intime du temps*. Paris: PUF.

Hwang, M. Y., Hong, J. C., Ye, J. H., Wu, Y. F., Tai, K. H., & Kiu, M. C. (2019). Practicing abductive reasoning: The correlations between cognitive factors and learning effects. *Computers & Education, 138*, 33-45.

Keats, J. (1993). *Lettres*. Paris: Belin.

Maturana, H., & Varela, F. J. (1994). *L'arbre de la connaissance*. Paris : Addison-Wesley.

Midtgarden, T. (2005). On The Prospects of A Semiotic Theory of Learning. *Educational Philosophy and Theory*, *37*(2), 239‑252. https://doi.org/10.1111/j.1469-5812.2005.00112.x

Midtgarden, T. (2010). Toward a Semiotic Theory of Learning. *Semiotics Education Experience*, 71-82. https://doi.org/10.1163/9789460912252_006

Oh, P. S. (2008). Adopting the abductive inquiry model (AIM) into undergraduate earth science laboratories. In *Science education in the 21st century* (pp. 263-277). Nova Science.

Oh, P. S. (2011). Characteristics of abductive inquiry in earth science : An undergraduate case study. *Science Education*, *95*(3), 409-430. https://doi.org/10.1002/sce.20424

Paavola, S., & Hakkarainen, K. (2005). Three Abductive Solutions to the Meno Paradox – with Instinct, Inference, and Distributed Cognition. *Studies in Philosophy and Education*, *24*(3), 235-253. https://doi.org/10.1007/s11217-005-3846-z

Peirce, C. S. (1868). Some consequences of four incapacities. *Journal of Speculative Philosophy,*

*2*, 140-157.

Peirce, C.S. (1994). *The collected paper of Charles Sanders Peirce* (Volumes I-VIII). Charlottesville: Intelex.

Pettersen, K. (2013). Acknowledging the role of abductive thinking: A way out of proceduralization for safety management and oversight? In C. Bieder & M. Bourrier (Eds.), *Trapping Safety into Rules* (pp. 107-117). Farnham: Ashgate.

Poizat, G., & Durand, M. (2015). Analyse de l'activité humaine et éducation des adultes : faits et valeurs dans un programme de recherche finalisée, *Revue française de pédagogie, 190*, 52-60.

Poizat, G., Durand, M., & Theureau, J. (2016). Challenges of activity analysis oriented towards professional training. *Le Travail Humain*, *79*, 233-258.

Poizat, G. & Flandin, S. (soumis). *Analyse de l'activité et formation : Vers un idéal de programme de recherche technologique*.

Poizat, G., & Goudeaux, A. (2014). Appropriation et individuation : un nouveau modèle pour penser l'éducation et la formation ? *TransFormations : Recherches en éducation et formation des adultes, 12*, 13-38.

Poizat, G., Salini, D., & Durand, M. (2013). Approche énactive de l'activité humaine, simplexité, et conception de formations professionnelles. *Education Sciences & Society, 4*, 97-112.

Récopé, M., Fache, H., Beaujouan, J., Coutarel, F., & Rix-Lièvre, G. (2019). A study of the individual activity of professional volleyball players: Situation assessment and sensemaking under time pressure. *Applied ergonomics*, *80*, 226-237.

Rosch, E. (1973). On the internal structure of perceptual and semantic categories. In T. Moore (Ed.), *Cognitive development and acquisition of language* (pp. 111-144). New York: Academic Press.

Rosenthal, V. (2012). La voix de l'intérieur. *Intellectica*, *58*(2), 53-89.

Romano, C. (1998). *L'événement et le monde*. Paris: PUF

Salini, D. (2013). *Inattendus et transformations de signification dans les situations d'information-conseil pour la validation des acquis de l'expérience*. Thèse non publiée pour le doctorat en sciences de l'éducation, Université de Genève.

Salini, D., & Durand, M. (2016). Événement dramatique et éducation événementielle, in J-P. Assal, M. Durand, & O. Horn (Eds.), *Le Théâtre du Vécu : Art, Soin, Éducation* (pp. 265-276). Dijon : Raison et passions.

Salini, D., & Durand, M. (2020). Overcoming a lived experience of personal collapse by creating a theatrical drama: an example of developmental and resilient adult education. In L. McKay, G. Barton, S. Garvis, & V. Sappa (Eds.), *Arts-based research, resilience and wellbeing across the lifespan* (pp. 169-189). Interdisciplinary handbook. Melbourne: Palgrave Macmillan.

Salini, D., & Poizat, G. (sous presse). Dénouements possibles de l'expérience d'impasse : pistes de compréhension et perspectives développementales. *Orientation scolaire et professionnelle. Dossier: L'expérience dans les pratiques d'accompagnement et de conseil en formation d'adultes*.

Schütz, A. (1987) *Le chercheur et le quotidien : Phénoménologie des sciences sociales*. Paris: Méridiens Klincksieck.

Simondon, G. (2005). *L'individuation à la lumière des notions de forme et d'information*. Grenoble: Millon.

Stables, A., & Gough, S. (2006). Toward a semiotic theory of choice and of learning. *Educational Theory*, *56*(3), 271-285.

Stiegler, B. (2010). *Ce qui fait que la vie vaut la peine d'être vécue : De la pharmacologie*. Paris: Flammarion.

Theureau, J. (2006). *Le cours d'action : Méthode développée*. Toulouse: Octarès.

Theureau, J. (2009). *Le cours d'action : Méthode réfléchie*. Toulouse: Octarès.

Theureau, J. (2010). Les entretiens d'autoconfrontation et de remise en situation par les traces matérielles et le programme de recherche « cours d'action ». *Revue d'anthropologie des connaissances, 4*(2), 287-322.

Theureau, J. (2015). *Le cours d'action : L'enaction et l'expérience*. Toulouse: Octarès

Varela, F. (1996). Neurophenomenology: A methodological remedy for the hard problem. *Journal of Consciousness Studies*, *3*(4), 330-349.

Ventimiglia, M. (2005). Three Educational Orientations : A Peircean Perspective on Education and the Growth of the Self. *Studies in Philosophy and Education*, *24*(3‑4), 291‑308. https://doi.org/10.1007/s11217-005-3851-2

Violi, P. (2014). *Paesaggi della memoria: Il trauma, lo spazio, la storia*. Torino: Bompiani.

Weick, K. (1993). The collapse of sensemaking: The Mann Gulch disaster. *Administrative Science Quarterly*, *38*(4), 62.

Weick, K. (2006). Faith, evidence, and action: Better guesses in an unknowable world. *Organization studies*, *27*(11), 1723-1736.

Weick, K., & Sutcliffe, K. (2001). *Managing the unexpected: Assuring high performance in an age of complexity*. San Francisco: Jossey-Bass.

Williams, M. (2018). *The aesthetic development: The poetic spirit of psychoanalysis - Essays on Bion, Meltzer, Keats* (2nd ed.). Abingdon: Routledge.

Zarifian, P. (1995). *Le travail et l'événement*. Paris: L'Harmattan.